\documentclass[referee,a4paper,12pt,traditabstract]{swsc} 

\usepackage{amsmath}
\usepackage{graphicx}
\usepackage{todonotes}
\usepackage{txfonts}
\usepackage{subfigure}
\usepackage{epstopdf}
\usepackage{lineno}
\usepackage[authoryear,round]{natbib}
\usepackage[backref]{hyperref}
\usepackage{url}
\usepackage{multirow}
\usepackage{graphicx}
\usepackage{arydshln}
\usepackage{makecell}

\graphicspath{{./figs/}}

\hypersetup{colorlinks=true,citecolor=cyan,urlcolor=cyan,linkcolor=blue}

\newcommand{\halpha}{H$\alpha$\,}
\newcommand{\halphanosp}{H$\alpha$}

\DeclareMathOperator*{\argmax}{arg\,max}

\usepackage{xcolor,colortbl}

\definecolor{Gray}{gray}{0.9}
\newcolumntype{a}{>{\columncolor{Gray}}c}
\definecolor{labels}{rgb}{0.133,0.54,0.133}

\begin{document}

\title{Real-time solar image classification: \\ assessing spectral,  pixel-based approaches}

\titlerunning{pixel-based solar image classification}

\authorrunning{Hughes et al.}

\author{J. Marcus Hughes \inst{1, 2} \and Vicki W. Hsu \inst{2, 3} \and Daniel B. Seaton \inst{2, 3} \and \\ Hazel M. Bain \inst{2,4} \and Jonathan M. Darnel \inst{2, 3} \and  Larisza Krista\inst{2, 3}}

\institute{
    Computer Science Department, University of Colorado Boulder, Boulder, Colorado, USA \\
    \email{\href{mailto:james.hughes-2@colorado.edu}{james.hughes-2@colorado.edu}}
        \and      
    Cooperative Institute for Research in Environmental Sciences, University of Colorado Boulder, Boulder, Colorado, USA
        \and
    National Centers for Environmental Information, National Oceanic and Atmospheric Administration, Boulder, Colorado, USA
    \and 
        Space Weather Prediction Center, National Oceanic and Atmospheric Administration, Boulder, Colorado, USA
        }

\abstract
{
In order to utilize solar imagery for real-time feature identification and large-scale data science investigations of solar structures, we need maps of the Sun where phenomena, or themes, are labeled. Since solar imagers produce observations every few minutes, it is not feasible to label all images by hand. Here, we compare three machine learning algorithms performing solar image classification using extreme ultraviolet and \halpha images: a {\color{black} maximum likelihood model} assuming a single normal probability distribution for each theme from Rigler et al. (2012), a maximum-likelihood model with an underlying Gaussian mixtures distribution, and a random forest model. We create a small database of expert-labeled maps to train and test these algorithms. Due to the ambiguity between the labels created by different experts, a collaborative labeling is used to include all inputs. We find the random forest algorithm performs the best amongst the three algorithms. The advantages of this algorithm are best highlighted in: comparison of outputs to hand-drawn maps; response to short-term variability; and tracking long-term changes on the Sun. Our work indicates that the next generation of solar image classification algorithms would benefit significantly from using spatial structure recognition, compared to only using spectral, pixel-by-pixel brightness distributions. 
}        


\maketitle

\section{Introduction}

{\color{black}At the National Oceanic and Atmospheric Adminstration's Space Weather Prediction Center (NOAA SWPC), forecasters utilize a variety of solar data, such as extreme ultraviolet (EUV) and soft-X-ray (SXR) images, X-ray irradiance measurements, and visible light images to draw labeled maps of the Sun twice each day. These maps indicate the locations of magnetic neutral lines, active regions, coronal hole boundaries, flares, and filaments and prominences. Such maps summarize the prevailing conditions on the Sun in a single view and can provide input to space weather models that can help forecast space conditions near Earth.}

However, transient solar events, such as flares, erupt and impact Earth over the course of minutes, requiring a map-making cadence faster than the twelve-hour cycle of current synoptic maps to capture these dynamics. Furthermore, the {\color{black} forecaster drawn maps} are available only in the portable document format (PDF) {\color{black} as drawn images and} cannot be easily used to identify regions of interest for scientific inquiry. 

A different approach is needed to complement the current method of solar drawings and address its limitations {\color{black} for SWPC usage}. Thus, in this paper, we compare {\color{black} a few} supervised machine learning approaches to generate solar thematic maps or images where the entire Sun is classified into different categories, called \textit{themes} in this paper. These maps will serve as the basis for a variety of new space weather products, that will be generated at SWPC in real-time (every four minutes) using solar imagery from the Solar Ultraviolet Imager \citep[SUVI,][]{Seaton2018} aboard NOAA's \textit{Geostationary Operational Environmental Satellite} (GOES) R-series satellites. 

There are many existing systems to identify regions of interest on the Sun \citep[see the review by][and references therein]{aschwanden2010}. Many of these utilize unsupervised machine learning (algorithms that learn from data that have no labels) and only classify a single theme. For example, \citet{curto+2008} used morphological transformations including erosion, dilation, opening, closing, and the top hat transformation to identify sunspot boundaries in \halpha images. \citet{benkhalil+2006} used Ca~\textsc{II}, \halphanosp, and extreme ultraviolet images (EUV) with a region growing technique to identify active region boundaries. \citet{kristagallagher2009} employed an EUV histogram thresholding method coupled with magnetograms to identify coronal holes and differentiate them from filaments.  Other classifiers also identify single themes, including: filaments \citep{qu+2005, fuller+2005, zharkova+2005}; active regions \citep{higgins+2011, mcateer+2005}; coronal mass ejections \citep{olmedo+2008}; sunspots \citep{preminger+2001, bratsolis+1998}; and flares \citep{borda+2002}. In addition, a suite of classifiers, many not listed here, were developed in association with NASA's \textit{Solar Dynamics Observatory} mission \citep{martens+2012}. With so many different classification approaches, \citet{cabellero+2013} conducted an independent study of {\color{black}a variety of approaches applied to active regions} and identified where they disagree. 

Unsupervised systems can also be multi-theme classifiers. For example, the Spatial Possibilistic Clustering Algorithm (SPoCA) is an unsupervised system that uses fuzzy logic to segment EUV solar images into regions of coronal hole, active region, and quiet sun. {\color{black} Both \citet{barra+2008} and \citet{verbeeck+2014} utilized this classifier to track active region and coronal hole evolution, with the latter paper describes some improvements and refinements to the original approach.}

These unsupervised techniques often perform well, but require extensive fine tuning and cannot be easily employed on user-defined categories. Supervised machine learning techniques utilize expert-labeled images to teach an algorithm how to properly identify themes. \citet{rigler+2012} proposed a maximum likelihood technique to produce real-time thematic maps from SUVI images based upon the earlier work of \citet{dewit2006}. This approach was to be used for operations, but operational limitations necessitate the use of a more robust classifier, such as those described in this paper. \citet{devisscher+2015} improved upon Rigler's approach by recognizing that the likelihood of different themes occurring in a single location was not the same for each theme and thus incorporated information about where these themes preferentially occur, i.e. spatial priors. That is to say, many coronal features, such as coronal holes, persist over long periods of time, and thus a pixel that was previously classified as a coronal hole is likely to remain classified as a coronal hole in future images. Non-probabilistic approaches, such as neural networks have also been applied to multi-theme solar feature classification \citep{kucuk+2017}.

In this paper we {\color{black}compare three methods for producing SUVI-derived thematic maps that have been implemented in the SUVI data processing software pipeline at SWPC.} We test the simple {\color{black} maximum likelihood model} of \citet{rigler+2012}, describe its operational limitations, and propose some refinements that can improve its accuracy, such as an underlying Gaussian mixtures model. Since this thematic map generator will be used in real-time for space weather operations, where data latency is a major concern \citep[see, for example, Figure 1 in][]{Redmon2018}, the model must be computationally efficient. To simplify the operational transition from \citet{rigler+2012}'s model to a new classifier, we currently constrain our study to approaches that adhere to the same classification strategy, namely classifying each pixel independently by only using data sources readily available on the operational systems at SWPC: the SUVI images, \halpha from GONG, and a derived, static limb model. We then compare our modified maximumum-likelihood approach to a random forest classifier and find that the random forest significantly outperforms all of the other classifiers in an operational setting. 

\section{Classifiers}
The techniques considered in this paper classify each image pixel independently based on its spectral properties, which is defined by the pixel intensity in each of the input images. Each pixel is classified into one of the following themes, $C$: quiet sun, bright region, coronal hole, flare, filament, prominence, and outer space. The input to the model consists of $q$ co-temporal input images, each consisting of $1280 \times 1280$ pixels in dimension, forming a $1280 \times 1280 \times q$ multi-channel image cube. A multi-channel pixel $\mathbf{x_{i}}$ is a vector $[x_{i,1}, x_{i,2}, \hdots, x_{i,q}]$, where $x_{i,n}$ is the value of the multi-channel image in the $i$-th pixel (imagine flattening the image into one row) in channel $n$. The output of the model is a new image, i.e. the thematic map at each time step, where each pixel within the thematic map has an associated output theme. {\color{black} Some approaches, such as the maximum likelihood and Gaussian mixtures techniques, fit the underlying probability distribution while other \textit{discriminative} approaches, such as random forests, do not directly fit the probability distribution.}

Three classifiers are tested: the model originally proposed by \citet{rigler+2012} ({\color{black}maximum likelihood} classifier with an assumed Gaussian distribution), a maximum-likelihood classifier with a fitted Gaussian mixtures model, and a random forest classifier. Each classifier takes a multi-channel image cube as input and outputs a thematic map. 

\subsection{{\color{black} Maximum likelihood} approach with normal distribution}
\citeauthor{rigler+2012} proposed the original {\color{black} maximum likelihood} method for creating thematic maps from SUVI data. They used a smoothness prior to ensure neighboring pixels share the same class and suppress spurious, small-scale class fluctuations within a given region. The version presented here disregards the smoothness prior because we expect the algorithm to perform well without additional filtering. The algorithm is based upon Bayes's theorem:

\begin{align}
P(c_m| \mathbf{x_{i}}) &= \frac{P(\mathbf{x_{i}}|c_m) P(c_m)}{P(\mathbf{x_{i})}}  
\end{align}
 
Thus, the probability for the $m$-th theme given multi-channel pixel $\mathbf{x_{i}}$, $P(c_m| \mathbf{x_{i}})$ is expressed in terms of quantities that can be calculated or disregarded. These quantities are the probability of the $m$-th theme, $P(c_m)$; the probability of the pixel given the $m$-th theme, $P(\mathbf{x_{i}}|c_m)$; and the probability of the pixel $P(\mathbf{x_{i}})$. Since the pixel probability, $P(\mathbf{x_{i}})$, is the same for each theme it does not impact which theme maximizes the likelihood and is disregarded. \citeauthor{rigler+2012} simplified the model further by disregarding the theme probability, $P(c_m)$, {\color{black} assuming there is no \textit{a priori} reason a pixel should have one theme instead of another and thus adopt the \textit{maximum likelihood estimation} model.

Here} maximizing $P(\mathbf{x_{i}}|c_m)$ will maximize $P(c_m | \mathbf{x_{i}})$. The inferred theme for a given pixel is then $\argmax_{c_m \in C} P(\mathbf{x_{i}}|c_m)$, that is the theme that has the highest probability. Using expert-labeled thematic maps, \citeauthor{rigler+2012} estimate a $k$-variate normal distribution to model the theme-conditioned probabilities. Each theme has $n_m$ pixels identified with it. The mean $\mu_m$ and covariance matrix $\Sigma_m$ for $m$-th theme are:
 
\begin{align}
\mu_m &= \frac{\sum_{i=1}^{n_m} \mathbf{x_{i,m}}}{n_m} \\
\Sigma_m &= \frac{\sum_{i=1}^{n_m} \left[ x_{i,m} - \mu_m \right] \times \left[x_{i,m} - \mu_m\right]^T}{n_m} \\
P(\mathbf{x_{i}} | c_m) &= \frac{\exp \left(-\frac{1}{2} \left(\mathbf{x_{i}} - \mu_m \right) ^ \mathrm{T} \Sigma_m^{-1} \left(\mathbf{x_{i}} - \mu_m \right) \right)}{\sqrt{\left(2 \pi\right)^k |\Sigma_m|}}
\end{align}

\subsection{Maximum-likelihood approach with Gaussian mixtures}
Upon further investigation, it was determined that the distributions of pixel intensities for each theme were not well approximated by a Gaussian distribution, see Section \ref{sec:lessons} for an in-depth discussion. To improve performance, the Gaussian model of \citeauthor{rigler+2012} for each theme was replaced with a Gaussian mixture. The Gaussian mixture is a weighted sum of $k$ Gaussian components each having a mean $\mu_i$ and covariance matrix $\Sigma_i$, fitted using the Expectation-Maximization algorithm \citep{dempster+1977}. In particular, we utilized the Gaussian mixture implementation from the Python package scikit-learn \citep{pedregosa+2011}. 

{\color{black}Other density estimation approaches allow further relaxation of the assumption that the distribution is fundamentally Gaussian in nature. For example, a non-parametric kernel density estimator might perform better than the Gaussian mixtures approach we describe here. However, a key goal of this paper is to characterize the performance of the maximum likelihood approach described in \citeauthor{rigler+2012} using real SUVI data, and thus we choose to consider only this small modification of the \citeauthor{rigler+2012}'s maximum likelihood classifier and not experiment with other density estimating approaches.}

\subsection{Random forest}
The aforementioned methods each assume something about the underlying theme spectral distribution, and classify each pixel by the distribution to which it is closest. However, if the spectral distribution is poorly fitted, which can happen for various reasons discussed further in Section~\ref{sec:lessons}, then the classification could also be poor. Furthermore, these distribution-based algorithms require evaluating a potentially complex and time-consuming probability distribution function. 

\begin{figure}
   	\centering
   	\includegraphics[width=\textwidth]{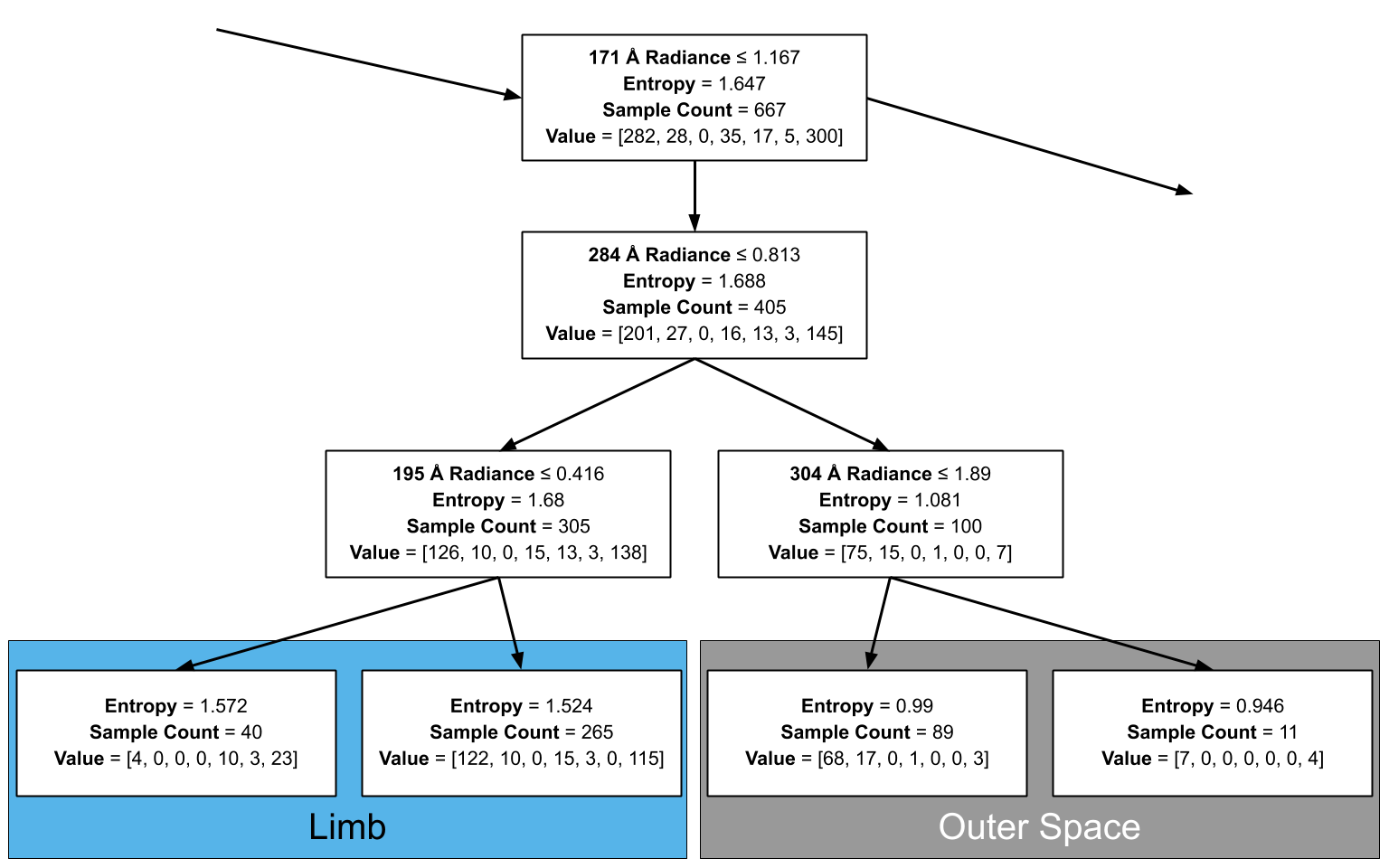}
   	\caption{\textbf{Random forest model:} A portion of an example decision tree. A multi-channel pixel moves down this tree until reaches a leaf node, here the colored boxes, and adopts that label. The blue leaf nodes were learned as limb while the gray leaf nodes are learned as outer space. }
   	\label{fig:decisiontree}
\end{figure}

An alternative, a random forest, utilizes an ensemble of decision trees, where each decision tree has a branching pattern of binary, yes-no decisions based on simple rules. These yes-no decisions occur at points in the tree called nodes. An example portion of a tree used in this application is shown in Figure~\ref{fig:decisiontree}. A multi-channel pixel moves from the top of the tree downward, until reaching a leaf node, or nodes in the tree where no further branching occurs. Each leaf node has an associated theme label which is used to classify the pixel. To move through the tree, at each internal node, a rule has been learned of whether the tree branches to the left or the right. In the example shown, the topmost node has the rule that the 171~\AA\ channel brightness is less than or equal to 1.167. If this condition is met for the pixel in question, the vector moves downward in the Figure. If the pixel is greater than 1.167 in 171~\AA\, the vector moves along the right arrow, out of this figure, into a separate portion of the tree that classifies when 171~\AA\ is brighter, potentially themes other than limb and outer space (not shown due to space limitations). {\color{black} The threshold value of 1.167 is learned during the training phase to optimize entropy, where entropy represents how pure the label distributions for a given leaf node are; the aim is to have a clear separation of labels when growing the trees.} In general, each decision tree partitions the input space completely into different themes. The random forest, being an ensemble of many trees, then robustly partitions the input space.

In order to best classify the training data, these branching rules are learned as described in \cite{Breiman2001}. Random forests utilize sampling with replacement, or ``bagging", a technique for iteratively selecting samples from a dataset where a copy is replaced in the original dataset before continuing. Each decision tree is trained on a bagged copy of the training data and thus are more robust to noise and the specific structure of the training database. Considering classifications from all of the decision trees, the random forest returns a more robust classification that is not as sensitive to these problems. The random forest is optimized by tuning several hyper-parameters, most importantly how many decision trees are used and how deep each tree can be. For this investigation, we used the Python scikit-learn implementation of random forests \citep{pedregosa+2011}. 

\section{Data sources} \label{sec:data}

Classified solar images are of particular value to space weather forecasters because solar regions --- particularly flares --- need to be identified as rapidly as possible in order to help them issue timely watches, warnings and alerts. \citep[See][for a discussion of the need for rapid response to changing solar conditions.]{Redmon2018} To achieve this, input data sources must be available with minimal latency, and thus we used data sources that are real-time, low-latency data streams available at NOAA for this study.

The EUV imagery we use comes from SUVI on board NOAA's GOES-16 and -17 spacecraft. EUV imagery in a variety of wavelengths provides a good overview of the prevailing solar conditions, but sometimes it is not sufficient to disambiguate between some themes, such as coronal holes and filaments. Thus, similarly to the approach discussed by \citet{DELOUILLE2018}, we supplement these observations with \halpha observations from the Global Oscillation Network Group (GONG) ground-based observatories, which provides additional information about chromospheric conditions.

\subsection{Extreme ultraviolet imagery}
SUVI is a normal-incidence EUV Ritchey-Chr\'etien telescope that observes the solar atmosphere in six ultraviolet wavelengths --- 93, 131, 171, 195, 284, and 304~\AA\ --- corresponding to emission from ion species at a range of temperatures between about 50,000 K and 20 MK. \citep[See the discussion in][for a summary of various ions and their emission spectra in EUV imagery.]{odwyer+2010} SUVI images are approximately $53.3 \times 53.3$~arcmin, $1280\times1280$~pixel square images with a plate scale of 2.5~arcsec.

Each SUVI instrument operates at a cadence of 10~s, obtaining images at a variety of wavelengths and exposure times, which allows software running on the ground to produce high dynamic range (HDR) composite images that can capture both quiet corona and bright solar flares in a single frame with no saturation. The effective HDR cadence varies between one and four minutes, depending on the passband. It is these HDR composite images that we use as inputs throughout this study.

\subsection{\halpha imagery}
GONG has produced full-disk H-alpha observations of the Sun since 2010 \citep{harvey+2011} using a ground-based network of six observatories, sufficiently spread across the planet to ensure continuous coverage at all times of day. {\color{black} GONG generates $2048 \times 2048$~pixel} images that cover the complete solar disk with 1~arcsec pixels. It is worth noting that because GONG is a ground-based network, image quality, pointing, and resolution can be influenced by local atmospheric conditions. GONG data is used for a variety of space weather applications.

Because the pointing, plate scales, and image roll of GONG and SUVI data are different, it is necessary to perform image scaling and alignment in order to use the data together. These transformations are handled partially by the pre-processing applied to SUVI data to generate HDR composites, and partly by additional steps during the analysis described here. These transformations are performed based on standard image navigation metadata in the files.

\subsection{Limb emission measure}\label{sec:limb}
In solar image classification, it is helpful to differentiate pixels on and off the solar disk since the differing line of sight length causes observational changes within the same structure class. This could be done explicitly by training separate on and off disk classifiers, but we follow the example of \citeauthor{rigler+2012} and incorporate a model of the solar corona as an input channel. This avoids agreement complications in labeling themes along the solar limb with different classifiers. We tested this approach using a simple, azimuthally symmetric model of the limb brightening effect derived from a semi-empirical model of coronal density described by \citet{sittlerguhathakurta1999}. {\color{black}However, our tests showed that the use of this synthetic channel did not improve the performance of the classifier both when considering EUV images alone and EUV and \halpha images together, and thus we did not use this synthetic channel in the analysis presented in this paper. With sufficient training data, it could be helpful. Instead, it became a confusing additional information that dominated the algorithms; they too strongly keyed on the limb emission measure. For example, we only have two flares, one in training and one in test, so the algorithm expects flares to have limb emission measure similar to the training example; we would need them distributed across the disk in training for this feature to be helpful. We added noise to the limb emission measure to decrease the dependence, but it was unsuccessful and we disregard this channel even though it is in the original \citeauthor{rigler+2012} model.}

\section{Annotated images}

\begin{table}[]
\centering
\begin{tabular}{|l|l|l|}
\hline
\textbf{Date} & \textbf{Type} & \textbf{Notes} \\ \hline
September 6, 2017 12:02:00 & Test & Includes X-class flare \\ \hline
September 10, 2017 16:06:00 & Train & Includes X-class flare \\ \hline
May 27, 2018 09:27:22 & Train &  \\ \hline
June 11, 2018 12:50:06 & Test &  \\ \hline
June 25, 2018 12:50:50 & Train & \\ \hline 
July 14, 2018 07:01:59 & Test &  \\ \hline
July 27, 2018 08:01:50 & Train & \\ \hline
October 9, 2018 00:01:00 & Test & \\ \hline
\end{tabular}
\caption{\textbf{Image dates:} Dates annotated by experts to use for training and performance evaluation. Train and test were alternated on a 14-day cycle so different sides of the Sun were observed, forcing independent sampling if the Sun had not changed considerably.}
\label{tab:dates}
\end{table}
 
Eight observations obtained at different times, listed in Table \ref{tab:dates}, were annotated by four solar physicists from NOAA and the University of Colorado Boulder. These dates were selected to include a variety of solar phenomena, especially bright flares. SWPC issues space weather alerts for flares of class M1.0 or higher, but none were observed in 2018 during normal operation of SUVI. Therefore, we used specially processed data from the bright flare events of September 2017, during SUVI's testing phase, to provide inputs to train the flare category \citep[see][for a discussion of these data]{Seaton2018}. The remaining observations contain a variety of bright regions, coronal holes, prominences, filaments, limb, and quiet sun. 
 
The experts were asked to annotate the images from the selected dates utilizing \href{https://github.com/jmbhughes/suvi-trainer}{custom software developed by author Hughes}. They independently assigned labels to pixels associated with identifiable themes on the Sun to determine their unbiased labeling. Upon meeting to discuss their labelings, it was clear there were many systematic differences, such as how large to draw the boundaries of bright regions. Consequently, the team {\color{black} discussed their independent (and generally subjective) criteria for labeling regions and agreed on consensus criteria that resulted in self-consistent labeling and boundaries throughout the observations and across thematic categories. For example, the experts agreed some individual labeled data overclassified flares to include most of the surrounding non-flaring active region, which led to poorly defined distributions in these categories, thus they arrived at a more conservative consensus definition on where to set the boundaries between these two types of features.} Experts then collectively re-labeled each image; each expert individually approved this new consensus labeling. When experts could not come to agreement or agreed the region contained artifacts that could not be labeled appropriately with any theme, the pixels were declared ``unlabeled." {\color{black} Only these consensus images were used in the machine algorithm.} The redundancy in expert labeling allows us to measure human solar image annotation performance using Fleiss's kappa, a metric for agreement among two or more annotators \citep{fleiss1973}:

\begin{equation} 
\label{eq:fleiss}
\kappa = \frac{\overline{P} - \overline{P_e}}{1-\overline{P_e}}.
\end{equation} 

The numerator of Equation \ref{eq:fleiss}, $\overline{P} - \overline{P_e}$, is the observed agreement above chance, and the denominator, $1 - \overline{P_e}$, corresponds to the maximal attainable agreement above chance, where random chance refers to all annotators randomly picking the same theme. These terms are calculated using equations \ref{eq:fleissexplain}--\ref{eq:fleissexplainend}, where $N$ is the number of pixels, $n$ is the number of ratings per pixel, and $M$ is the number of classes each pixel can be assigned. Each pixel is index $i = 1, 2, ..., N$ and each class is indexed {\color{black}$m = 1, 2, ..., M$}. Then, $n_{im}$ is the number of annotators who labeled the $i$-th pixel with class $m$. 

\begin{align}
p_m &= \frac{1}{Nn} \sum_{i=1}^{N} n_{im} \label{eq:fleissexplain} \\
P_i &= \frac{1}{n(n-1)} \left(\sum_{m=1}^{M} n_{im}^2 - n_{im} \right) \\
\overline{P} &= \frac{1}{N} \sum_{i=1}^N P_i \\
\overline{P_e} &= \sum_{m=1}^M p_m^2 \label{eq:fleissexplainend}
\end{align}

The scale for Fleiss's kappa shown in Table \ref{tab:fleiss} was proposed by \citet{landiskoch1977} to interpret metric results.

\begin{table}[ht!]
	\centering
	\begin{tabular}{|c c|}
		\hline    
		$\kappa$ range &  Interpretation \\
		\hline \hline
		$<$ 0 & No agreement \\ 
		0.0 - 0.19 & Poor agreement \\ 
		0.20 - 0.39 & Fair agreement \\
		0.40 - 0.59 & Moderate agreement \\
		0.60 - 0.79 & Substantial agreement \\ 
		0.80 - 1.00 & Almost perfect agreement \\
		\hline
	\end{tabular}
	\caption{\textbf{Interpretation of Fleiss's kappa coefficient for agreement:} from \citet{landiskoch1977}.}
	\label{tab:fleiss}
\end{table}

Using all themes and training data, we achieved a Fleiss's $\kappa$ of 0.94, which would indicate a remarkably high level of agreement. However, this is inflated because users were prompted with a solar disk pre-labeled with regions expected to be present over much of the image, either ``quiet sun'', ``limb'', and ``outer space,'' as appropriate to the particular location in the image. Since these themes account for the majority of the pixels in the image, it inflates the Fleiss's kappa value. If we only consider agreement on coronal hole, bright region, filament, prominence, and flare we get a $\kappa=0.38$, indicating only fair agreement. 

\begin{figure}
   	\centering
   	\includegraphics[width=0.5\textwidth]{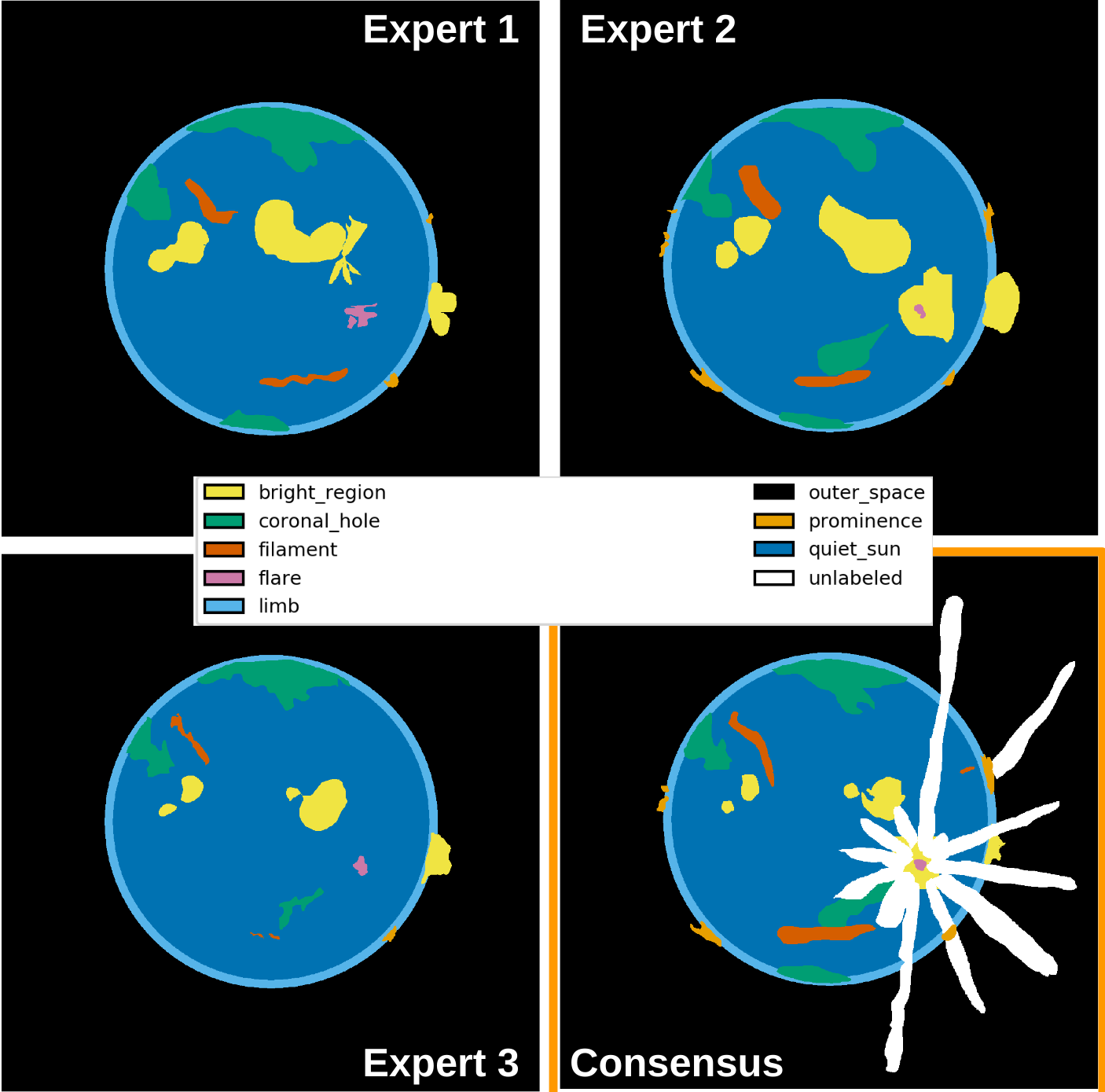}
   	\caption{\textbf{Expert inconsistency:} Illustration of identification inconsistencies between three different users for the 2018-09-06 12:02:00 observation. Three individually drawn expert labelings are shown with the collaborative consensus map in the bottom right. Note that the spikes (a result from diffraction due to the metallic mesh that provides support for SUVI's thin foil filters) from the bright flare are termed as ``unlabeled'' to avoid using the corresponding pixels in training or analysis.}
   	\label{fig:ExpertAgreement}
\end{figure}

When shown the same images, different experts arrive at different labeling conclusions for different solar phenomena. This disagreement comes from many sources: imprecision in hand-drawing the boundaries of the various labels; human subjectivity in determining where to draw boundaries led to different sized regions in different experts' labeling; experts missing faint phenomena like weak filaments habitually; and, occasionally, experts making clear errors in labeling. An example of this is shown in Figure \ref{fig:ExpertAgreement}.

{\color{black} The question of how to update the training data that drives these algorithms to account for changes in both instrument performance and overall changes in the character of the Sun as the solar cycle unfolds is an important one, since the thematic map will be used operationally and needs to provide consistent and reliable output over the long term. Our team's intent is to leverage ongoing workflows used by forecasters --- for example, solar drawings --- to help provide new expert-labeled data for re-training. There is little evidence for a need to retrain on timescales of weeks, and we currently lack sufficient data (that is, enough evolution in conditions on the Sun) to assess the longer-term need. Regardless, how to implement specific strategies to ensure operational consistency is a decision that belongs strictly to the forecast team at SWPC, and is beyond the scope of this paper. However, we emphasize the importance of good decision-making in this respect and will work, as appropriate, with SWPC forecasters to assist in developing strategies to ensure the long-term performance of the Thematic Maps classifiers they implement.
}

{\color{black}
\section{Model fitting}
The data was collected as discussed in Section \ref{sec:data}. Only the consensus expert images were used; individual labeling was not used in training or evaluation. The data was separated into training and testing according to Table \ref{tab:dates}. The train/test labels alternate every two weeks to ensure that train and test images represent statistically independent sample images, due to solar rotation. Thus, test and train images represent independent samples. For each test and train image all the pixels associated with ``unlabeled" regions, where experts could not reach consensus, were discarded and visual quality inspection was done for all images. 

Hyper-parameter estimation was completed with a grid search using $k$-fold validation on the training set. The simple Gaussian model lacks any hyper-parameters. For the Gaussian mixtures model, the number of components, $k$,  was assumed constant for each theme. A 50-fold validation experiment was conducted testing all odd number of components between 1 and 99, inclusive. Too many components gradually decreased the accuracy, by around 10\%, likely because the abundance of components was difficult to fit. For both the EUV and \halpha experiments in Table \ref{tab:perftab}, 20 components were used to optimize the trade-off between accuracy and run-times. 

The random forest model has more hyper-parameters to tune, some are redundant or known to not greatly impact training. We focused on fitting the number of trees in the forest, the maximum depth a forest can obtain, and the minimum number of samples for a leaf node. The number of trees is an insensitive parameter; more trees will not lead to overfitting but performance benefits will plateau after a critical number of trees has been reached \cite{Breiman2001}. We found that 250 trees was sufficient. Maximum depth and minimum number of samples are related parameters. Too shallow a tree forces a high minimum number of samples per leaf node and decreases the classification performance of a given tree since it cannot learn sufficient representation. An unbounded depth for a tree would result in nodes with single samples and are more sensitive to random sampling of the data. They are also more time and space expensive since classifying a new pixel requires many more evaluations which must be stored. We found that requiring 5 samples per leaf node and a maximum depth of 25 provide the best performance in validation tests using a grid search over depth limits from 1 to 100 depth at steps of 5, minimum leaf samples from 1 to 100 samples at steps of 5, and number of trees logarithmically gridded  40 times from 1 to 1000 trees. 
}
\section{Algorithm Analysis}
We evaluate the performance of our classifiers in three categories as proposed by \citet{devisscher+2015} {\color{black} using hold-out test data never used in training or hyper-parameter estimation}: through comparison to human annotated maps; short-term temporal stability; and long-term temporal stability. 
 
\subsection{Comparison to human maps}
Classifier-predicted pixel labels in a usable classification tool must match the expert consensus testing classifier values. We quantify this agreement in the standard machine learning metrics based on true positive count (TP), true negative count (TN), false positive count (FP), and false negative count (FN). For a given theme label, TP is the number of pixels classified with a given class for which the gold standard database agreed, {\color{black} i.e. the expert consensus database}. On the other hand, FP is the number of pixels classified with a given class for which the gold standard database disagreed. Similarly, TN is number of times the classifier \textit{correctly} did not assign the class and FN refers to cases where the classifier labeled a class other than the expert-designated class. We consider the following metrics, which are defined as:

 \begin{align}
     \mathrm{accuracy} &= \frac{TP + TN}{TP +  TN + FP + FN}, \\
     \mathrm{precision} &= \frac{TP}{TP + FP}, \\
     \mathrm{recall} &= \frac{TP}{TP + FN}, \\
     \mathrm{f1-measure} &= \frac{2 \times \mathrm{precision} \times \mathrm{recall}}{\mathrm{precision} + \mathrm{recall}}, \\
     {\label{eq:TSS}\color{black} \mathrm{TSS}} &{= \color{black} \frac{TP}{TP + FN} + \frac{TN}{TN+FP} - 1 }.
 \end{align}

{\color{black}Where TSS in Equation~\ref{eq:TSS} above refers to the ``true skill statistic'', which provides a performance metric that can characterize performance across classes of varying sizes without bias stemming from the sample ratios \citep[see the discussion in][]{bloomfield+2012}.}

 \begin{table}[]
 \centering
 \caption{\textbf{Algorithm performance when comparing to expert maps:} Accuracy, precision, recall,{\color{black} f1-measure and true skill statistic (TSS)} for each classifier. The first column under each metric shows scores when only EUV images were used, while the second column shows scores when \halpha was included. ``G" denotes the Gaussian model of \citet{rigler+2012}; ``M" is our Gaussian mixtures model; ``F" is our random forest model. At the bottom of the table, the average row shows the overall performance, with each class equally weighted. The entries highlighted green indicate which classifier and image set combination performed best on that metric.}
 
 \begin{tabular}{ll|aa|ll|aa|ll|aa}
 &  & \multicolumn{2}{c|}{Accuracy} & \multicolumn{2}{c|}{Precision} & \multicolumn{2}{c|}{Recall} & \multicolumn{2}{c|}{f1-measure} & \multicolumn{2}{c}{TSS} \\
& & EUV & \halpha & EUV & \halpha & EUV & \halpha & EUV & \halpha & EUV & \halpha \\
\hline
\multirow{3}{*}{outer space}
& G & 0.97 & 0.96 & {\color{labels}0.99} & {\color{labels}0.99}  & 0.96 & 0.96 & 0.97 & 0.97 & 0.94 & 0.93 \\
& M & {\color{labels}0.98} & 0.97 & 0.98 & 0.98  & 0.98 & 0.98 & 0.98 & 0.98 & {\color{labels}0.95} & 0.94 \\
& F & {\color{labels}0.98} & {\color{labels}0.98} & 0.98 & 0.98  & {\color{labels}0.99} & 0.98 & {\color{labels}0.99} & 0.98 & {\color{labels}0.95} & 0.94 \\
\hline
\multirow{3}{*}{bright region}
& G & {\color{labels}0.99} & {\color{labels}0.99} & 0.31 & 0.31  & {\color{labels}0.73} & {\color{labels}0.73} & 0.43 & 0.44 & {\color{labels}0.72} & {\color{labels}0.72} \\
& M & {\color{labels}0.99} & {\color{labels}0.99} & 0.36 & 0.38  & 0.69 & 0.69 & 0.47 & 0.49 & 0.68 & 0.68 \\
& F & {\color{labels}0.99} & {\color{labels}0.99} & 0.58 & {\color{labels}0.61}  & 0.53 & 0.55 & 0.55 & {\color{labels}0.58} & 0.53 & 0.55 \\
\hline
\multirow{3}{*}{filament}
& G & 0.96 & 0.96 & 0.03 & 0.04  & 0.51 & {\color{labels}0.54} & 0.06 & 0.08 & 0.47 & {\color{labels}0.51} \\
& M & 0.97 & 0.97 & 0.04 & 0.04  & 0.46 & 0.42 & 0.07 & 0.08 & 0.43 & 0.40 \\
& F & {\color{labels}0.99} & {\color{labels}0.99} & {\color{labels}0.06} & {\color{labels}0.06}  & 0.21 & 0.30 & {\color{labels}0.10} & {\color{labels}0.10} & 0.20 & 0.28 \\
\hline
\multirow{3}{*}{prominence}
& G & 0.97 & 0.98 & 0.06 & 0.08  & 0.63 & {\color{labels}0.79} & 0.11 & 0.15 & 0.60 & {\color{labels}0.77} \\
& M & {\color{labels}0.99} & {\color{labels}0.99} & 0.12 & {\color{labels}0.21}  & 0.54 & 0.60 & 0.19 & {\color{labels}0.31} & 0.53 & 0.59 \\
& F & {\color{labels}0.99} & {\color{labels}0.99} & 0.17 & 0.17  & 0.39 & 0.56 & 0.23 & 0.26 & 0.39 & 0.55 \\
\hline
\multirow{3}{*}{coronal hole}
& G & 0.92 & 0.95 & 0.16 & 0.22  & {\color{labels}0.91} & {\color{labels}0.91} & 0.27 & 0.35 & 0.83 & 0.86 \\
& M & {\color{labels}0.98} & {\color{labels}0.98} & 0.42 & 0.43  & 0.86 & 0.90 & 0.56 & 0.58 & 0.84 & 0.88 \\
& F & {\color{labels}0.98} & {\color{labels}0.98} & {\color{labels}0.51} & 0.45  & 0.85 & {\color{labels}0.91} & {\color{labels}0.64} & 0.60 & 0.84 & {\color{labels}0.90} \\
\hline
\multirow{3}{*}{quiet sun}
& G & 0.87 & 0.90 & 0.89 & {\color{labels}0.93}  & 0.51 & 0.61 & 0.65 & 0.74 & 0.50 & 0.60 \\
& M & 0.93 & 0.94 & 0.90 & {\color{labels}0.93}  & 0.77 & 0.79 & 0.83 & 0.85 & 0.74 & 0.77 \\
& F & {\color{labels}0.96} & 0.95 & 0.91 & 0.92  & {\color{labels}0.90} & 0.86 & {\color{labels}0.90} & 0.89 & {\color{labels}0.87} & 0.84 \\
\hline
\multirow{3}{*}{limb}
& G & 0.97 & 0.96 & 0.39 & 0.31  & 0.52 & 0.60 & 0.45 & 0.41 & 0.50 & 0.57 \\
& M & {\color{labels}0.98} & 0.97 & 0.47 & 0.42  & 0.78 & 0.83 & 0.59 & 0.56 & 0.76 & 0.81 \\
& F & {\color{labels}0.98} & {\color{labels}0.98} & {\color{labels}0.51} & 0.48  & 0.84 & {\color{labels}0.89} & {\color{labels}0.64} & 0.62 & 0.82 & {\color{labels}0.87} \\
\hline
\multirow{3}{*}{flare}
& G & {\color{labels}1.00} & {\color{labels}1.00} & 0.07 & 0.07  & 0.97 & 0.97 & 0.12 & 0.12 & 0.97 & 0.97 \\
& M & {\color{labels}1.00} & {\color{labels}1.00} & 0.06 & 0.02  & {\color{labels}1.00} & {\color{labels}1.00} & 0.11 & 0.04 & {\color{labels}1.00} & {\color{labels}1.00} \\
& F & {\color{labels}1.00} & {\color{labels}1.00} & 0.40 & {\color{labels}1.00}  & 0.99 & 0.08 & {\color{labels}0.57} & 0.15 & 0.99 & 0.08 \\
\hline

\hline \hline
\Xhline{2\arrayrulewidth}

\multirow{3}{*}{\textbf{average}}
& \textbf{G} & 0.96 & 0.96 & 0.36 & 0.37 & 0.72 & 0.76 & 0.38 & 0.41 & 0.69 & 0.74 \\
& \textbf{M} & {\color{labels}0.98} & {\color{labels}0.98} & 0.42 & 0.43 & 0.76 & {\color{labels}0.78} & 0.48 & 0.49 & 0.74 & {\color{labels}0.76} \\
& \textbf{F} & {\color{labels}0.98} & {\color{labels}0.98} & 0.52 & {\color{labels}0.58} & 0.71 & 0.64 & {\color{labels}0.58} & 0.52 & 0.70 & 0.63 
\end{tabular}%

\label{tab:perftab}
\end{table}

\begin{figure}
   	\centering
   	\includegraphics[width=0.5\textwidth]{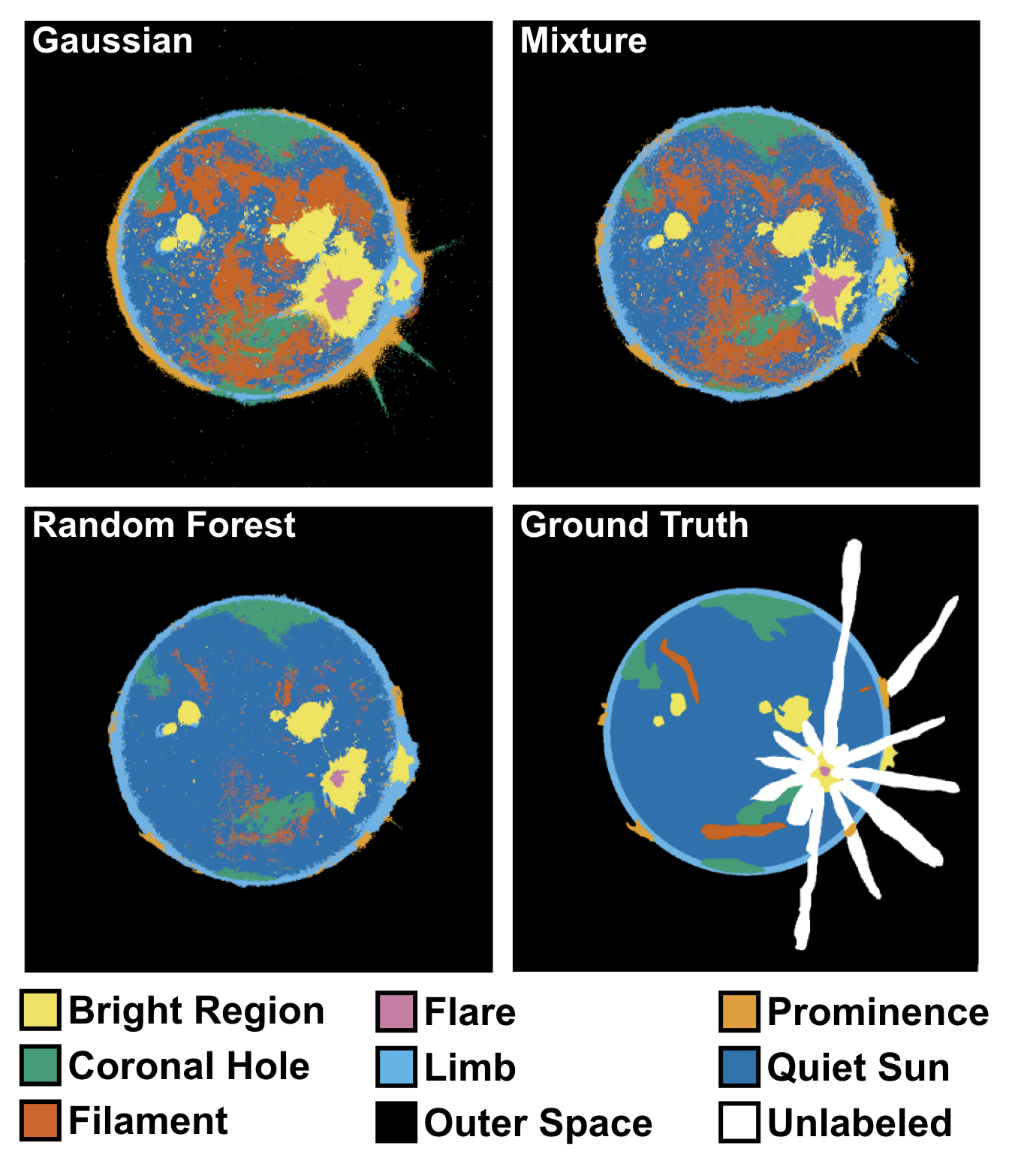}
   	\caption{Algorithm test images from the 2017~Sep~6 X9.3 flare using only EUV channels as input. Filament classification improves (see Figure~\ref{fig:comparison_halpha}) when adding \halpha.}
   	\label{fig:comparison_euv}
\end{figure}

\begin{figure}
   	\centering
   	\includegraphics[width=0.5\textwidth]{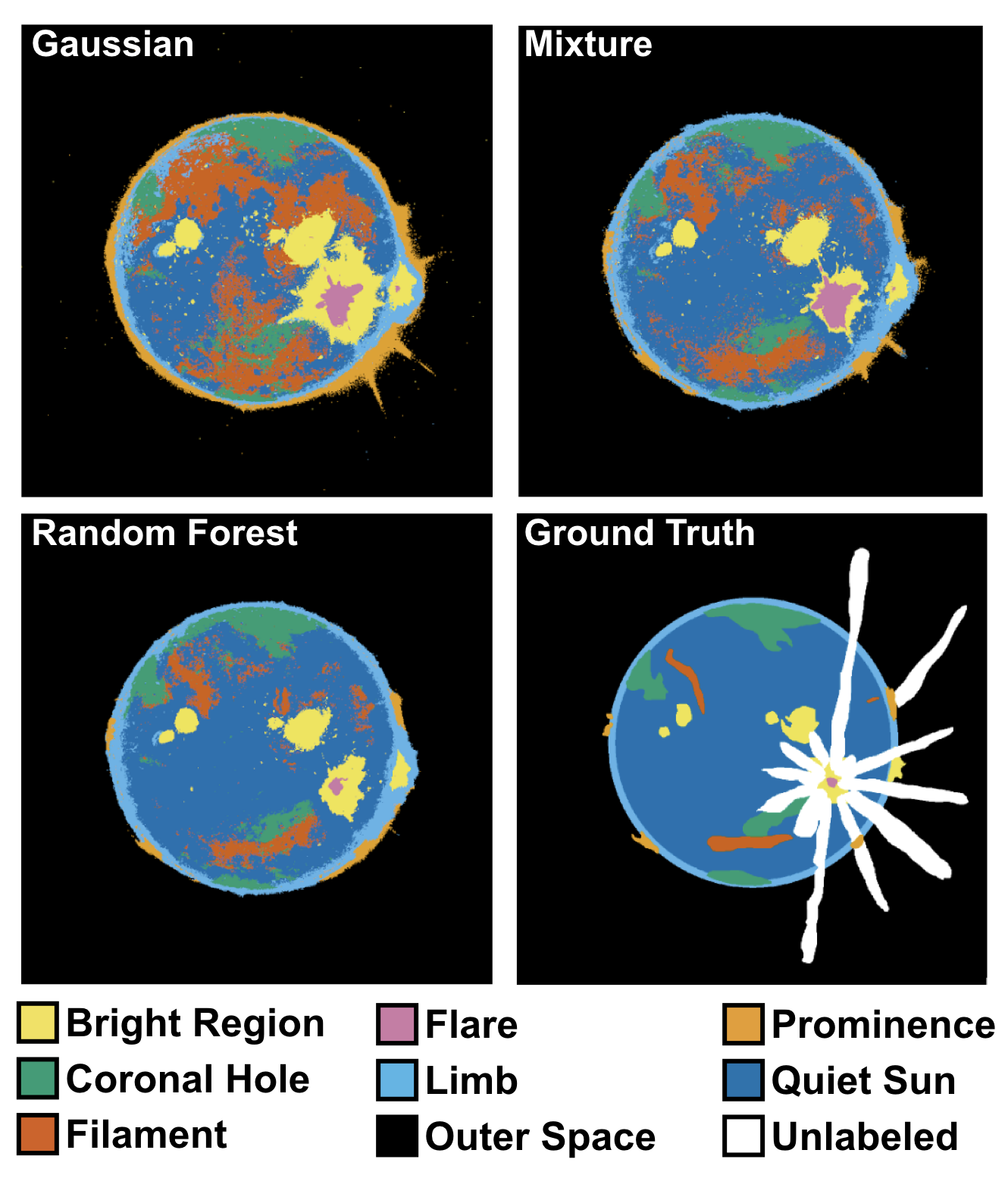}
   	\caption{Algorithm test images from the 2017~Sep~6 X9.3 flare using both EUV and \halpha channels as input.}
   	\label{fig:comparison_halpha}
\end{figure}

 The results are summarized in Table \ref{tab:perftab}. The main observation is that, quantitatively, the random forest algorithm performs the best by maximizing the f1-measure, {\color{black} a metric that is robust to class imbalance and balances true positive rates with errors,} of most themes. Figures~\ref{fig:comparison_euv} and \ref{fig:comparison_halpha} qualitatively confirm this conclusion and provide spatial insight on how and where each classifier tends to fail, an important characteristic not directly observed in the performance metrics. The Gaussian and mixture models over-identify the filament theme, and their flare boundary is much larger than the random forest. The different approaches would likely converge on flares with more training examples. This would more completely sample the underlying theme distributions with dimmer flares to distinguish from bright regions. 
 
 The addition of the \halpha channel in Figure \ref{fig:comparison_halpha} greatly constrains the filament and prominence classes. Before including the \halpha channel, the random forest indicated scattered filament labeling, but after including \halpha, the classifier becomes more confident. This trend is seen in all the test images. 

\begin{figure}
   	\centering
   	\includegraphics[width=0.5\textwidth]{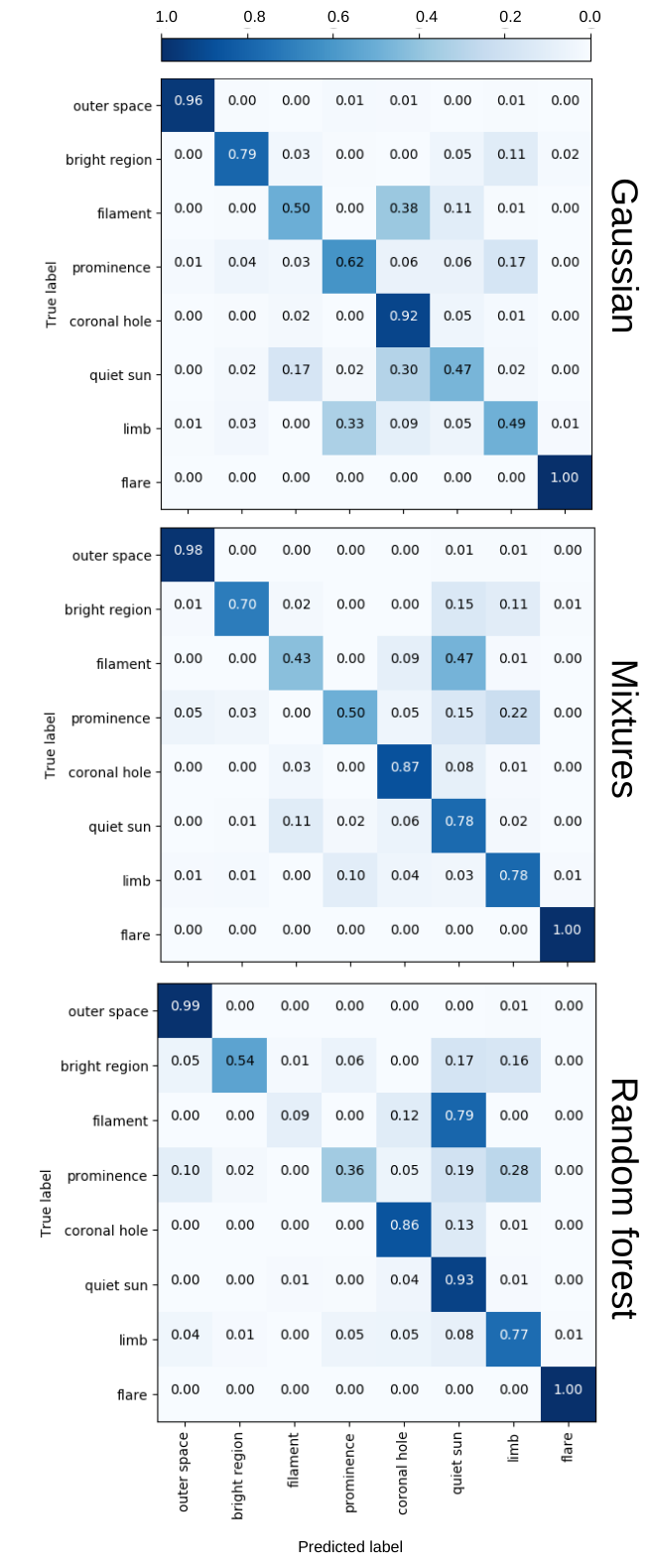}
   	\caption{Algorithm confusion matrices for classifiers with \halpha. Darker blue indicates a more perfect classification while light blue indicates no pixels fall in this category.}
   	\label{fig:cm}
\end{figure}

To assess which themes the algorithms systematically confuse (for example, flare versus bright region and filament versus quiet sun), we generated confusion matrices in Figure \ref{fig:cm} for classifiers with the \halpha channel included. A confusion matrix compares the theme labels identified by automated tools with those identified in the expert-classified reference set. For an approach that performs perfectly, the diagonals would all be ones, indicating perfect labeling. Off diagonal entries indicate which themes are misidentified, and how such themes are classified. 

It is interesting to note that the performance of each algorithm in the filament classification worsens from the Gaussian approach to the random forest approach according to some metrics. In fact, this is because the classifier becomes --- appropriately --- more conservative about applying the filament theme. The expert-labeled boundaries were generous and identified some filaments that were not obvious or, in other cases, applied the label to too broad a region. Such areas are classified as quiet Sun by the random forest tool. This creates the appearance of a disagreement in labeling that could be considered poor performance, when, in fact, such conservative behavior is likely desirable.

{\color{black}This behavior is a good reminder that it requires a variety of metrics --- and an understanding of the context in which these algorithms are applied --- to fully evaluate the true performance of any particular classifier. For example, as we note, a conservative classifier might generate no false positives but many false negatives compared to our relatively coarse expert training set. Even though we might strongly prefer this behavior to having numerous false detections of features of space weather significance, some metrics, like the TSS, give a poor score to a classifier that behaves in this way. One must therefore consider the classifier's performance relative to a variety of metrics with differing characteristics and a broad view of what behavior one considers to be desirable in such a tool in order to fully evaluate the success of the classifier.

On the other hand, metrics such as these can help shine a light on important limitations of various approaches. For example, the TSS makes evident that the addition of \halpha to the random forest yields a poor flare performance, which is not as evident from some of the other performance measures. This is likely because flares near the limb where the footpoints are not visible might not have a clear \halpha signature, while on-disk flares do. The mixing of these two data can create ambiguous signals in the classifier and lead to false negatives, to which the TSS is especially sensitive.}

\subsection{Short-term temporal instability}
In general, the Sun changes little within four minutes, so a thematic map and its subsequent map four minutes later should be nearly identical. Except in the case of highly dynamic phenomena, if two such maps are not identical, this indicates classifier is likely not robust to noise. To assess this, we examined how many pixels changed classification from one map to the next. Given a thematic map, $A$, and a subsequent thematic map 4 minutes later, $B$, we define the short term {\color{black} instability} as the percentage of pixels that change classification (given that SUVI images are 1280$\times$1280 pixels):

\begin{align}
    \text{short-term temporal {\color{black} in}stability} &= \frac{\sum_{i=1}^{1280} \sum_{j=1}^{1280} \delta (A_{i,j}, B_{i,j})}{1280 \times 1280} \\
    \delta(t_1, t_2) &=  \begin{cases} 
                1 & t_1 \neq t_2 \\
                0 & \mathrm{otherwise} 
  \end{cases}
\end{align}
 
\begin{figure}
   	\centering
   	\includegraphics[width=\textwidth]{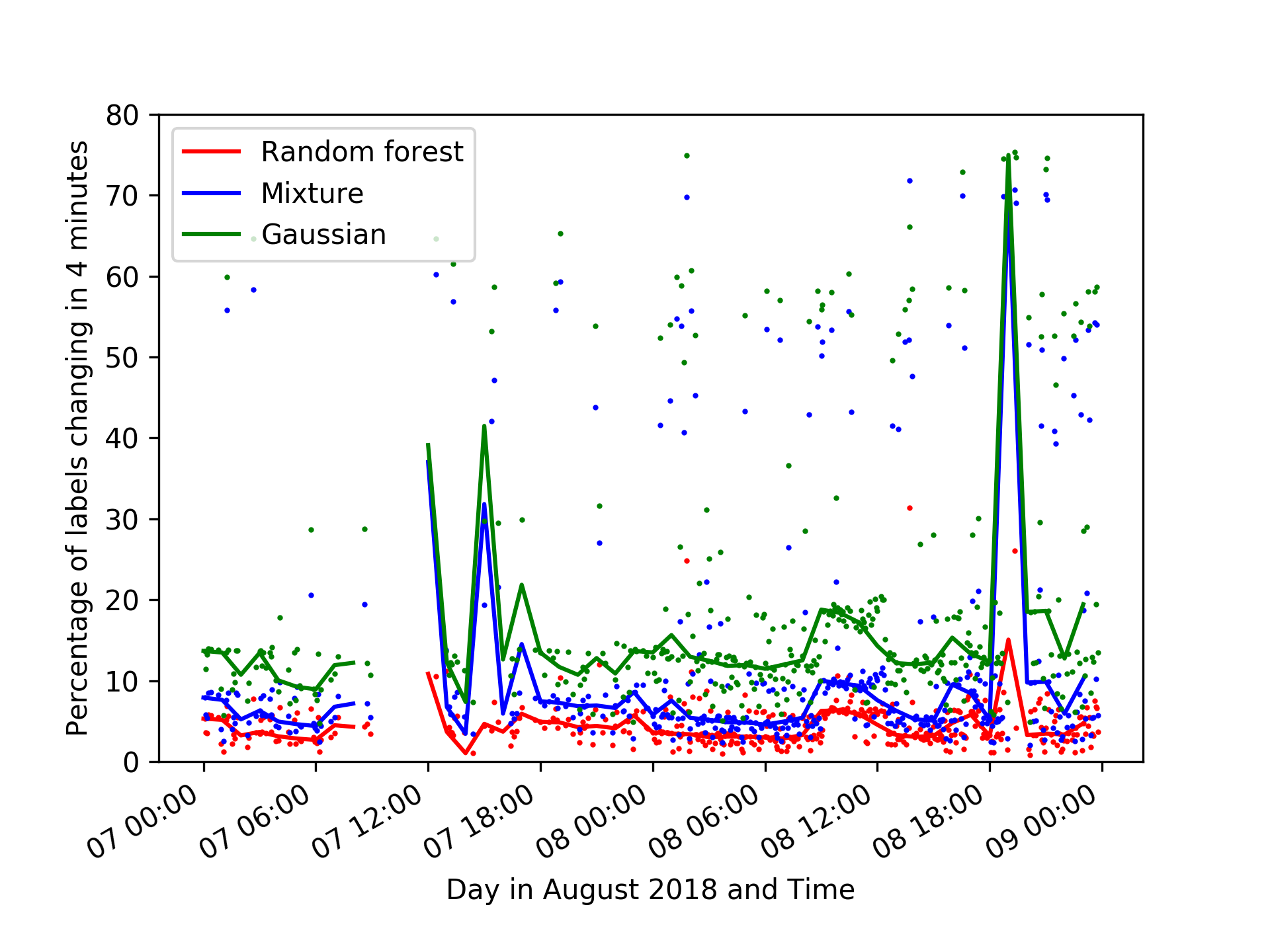}
   	\caption{\textbf{Short term stability metric:} Rate of change in pixel labels over time for each classifier using the EUV and \halpha images.}
   	\label{fig:TemporalStability}
\end{figure}

\begin{figure}
   	\centering
   	\includegraphics[width=\textwidth]{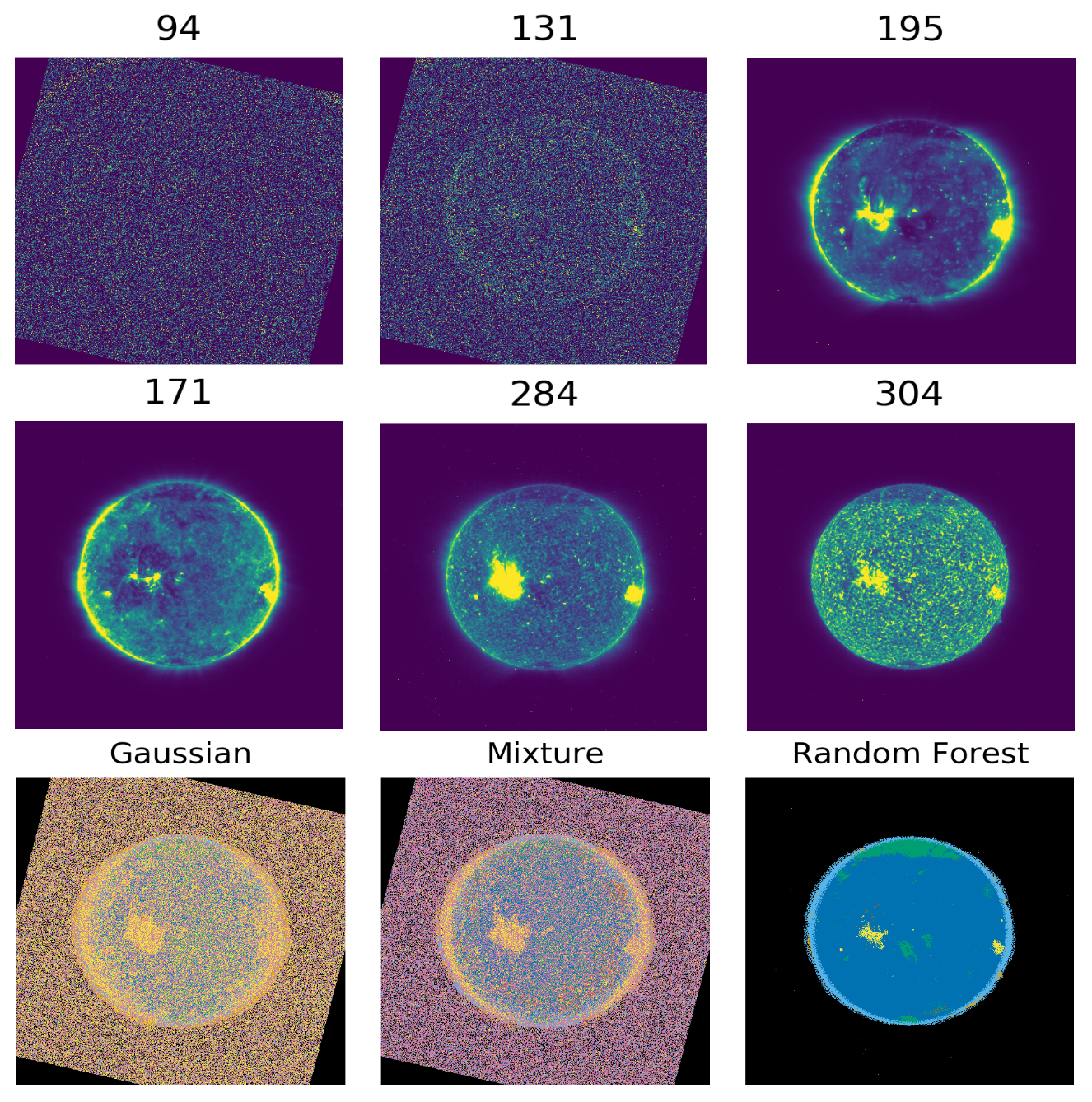}
   	\caption{\textbf{Robustness of random forest:} Input images for each EUV channel, scaled such that the bottom and top 1\% of pixel values are clipped (top two rows). The 94~\AA\ and 131~\AA\ images here have only the short exposure images and barely show the Sun. The single Gaussian and mixture approaches fail to create good thematic maps while the random forest approach is more robust to errors (bottom row).}
   	\label{fig:missing}
\end{figure}

As shown in Figure~\ref{fig:TemporalStability}, the random forest is more robust to noise than the mixtures or single Gaussian model. The random forest had a 4.4\%/3.6\% mean/median percentage of pixels change label over the tested {\color{black} two-day time period (about 720 image pairs)} while the mixtures approach had 13.1\%/6.3\%, and the single Gaussian had 19.5\%/13.1\% change. Even in the event when high-dynamic range images for some channels are severely degraded because of a missing input image, as shown in Figure \ref{fig:missing}, the random forest was able to create viable thematic maps, while the other approaches could not. For pixels where some channel values are erroneous due to image artifacts, the random forest can recover the correct answer based on the decisions on valid pixel values in other channels. In the maximum likelihood approaches, such a pixel ends up in a portion of the parameter space never seen in training, and since the decision is a single calculation, the classifier cannot recover the correct label. Figure \ref{fig:TemporalStability} confirms this because the random forest change is always less than the other approaches and rarely spikes above 10\%. We found no previous quantitative analysis of the stability of solar image segmentation algorithms in literature to which we could compare.
 
\subsection{Long-term temporal stability}

\begin{figure}
   	\centering
   	\includegraphics[width=\textwidth]{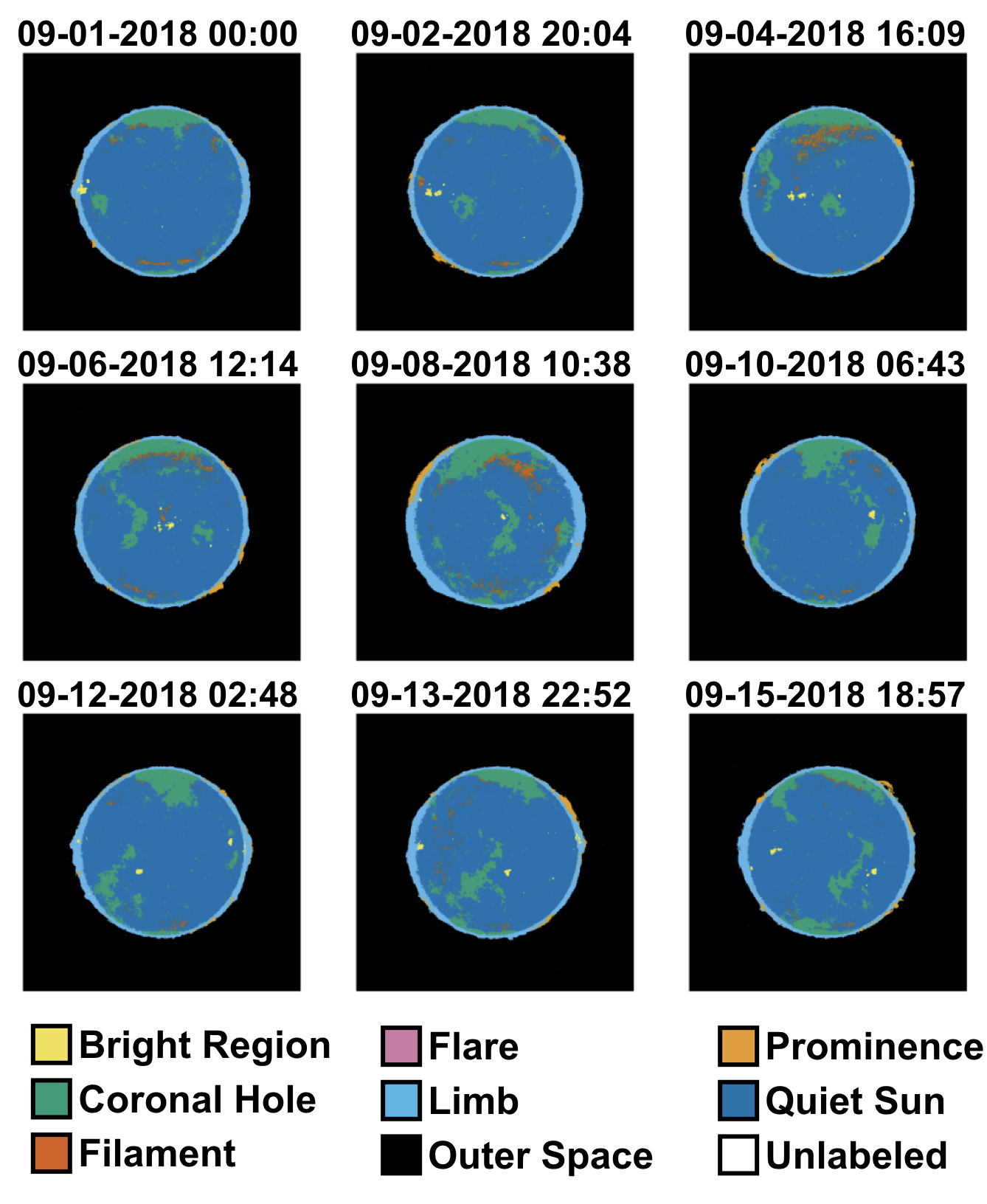}
   	\caption{\textbf{Long term stability metric:} Example thematic maps generated by the random forest algorithm with \halpha showing stable tracking of features across the Sun.}
   	\label{fig:longchange}
\end{figure}

Figure \ref{fig:longchange} shows the random forest's performance in segmenting the Sun over a duration of {\color{black}15 days, chosen because it allowed us to track the complete trajectory of a coronal hole and active region that were visible near the east limb at the beginning of the period of interest}. \halpha images are helpful in improving the labeling, however clouds that occasionally appear in ground-based \halpha observations can lead to artifacts, and thus more instability in labeling. Identifying when clouds are present in \halpha images and finding an appropriate recourse --- e.g. using a classifier without \halpha inputs or finding nearly co-temporal, cloud-free image --- is beyond the scope of this paper. 

\subsection{Comparison to SPoCA}
\begin{figure}
   	\centering
   	\includegraphics[width=0.5\textwidth]{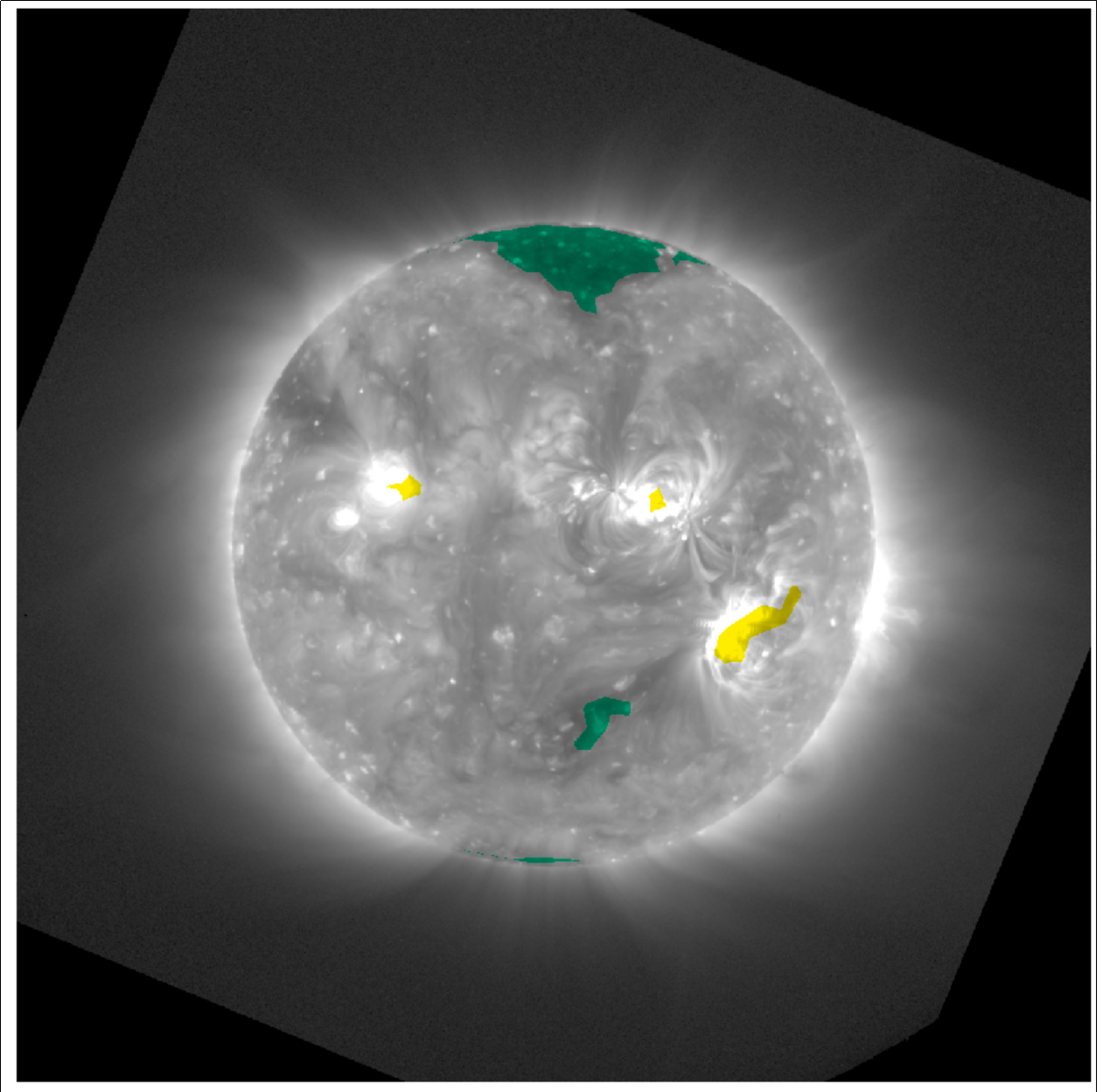}
   	\caption{\textbf{Comparison to SPoCA:} {\color{black} SPoCA classification for September 6, 2017 12:02:00, the same observation used in Figures~\ref{fig:ExpertAgreement}, \ref{fig:comparison_euv}, and \ref{fig:comparison_halpha}, overlaid on a contemporaneous SUVI 195~\AA\ image. SPoCA labels follow the thematic map color scheme: yellow for bright region and green for coronal hole.} }
   	\label{fig:spoca}
\end{figure}

SPoCA is a commonly used solar image segmentation system that outputs boundaries for active regions and coronal holes \citep{verbeeck+2014}. It runs constantly with results output to the Heliophysics Event Knowledgebase (HEK) \citep{hek}. We pulled the classifications for test observations {\color{black} and qualitatively compared its results to ours}. The random forest {\color{black} performs similarly to SPoCA} as seen in Figure \ref{fig:spoca}. {\color{black} SPoCA tended to be more conservative than the expert labeling. It also excluded one of the coronal holes our experts labeled and our algorithms identified. We did not include a quantitative comparison because the SPoCA boundaries available from HEK are different from the true SPoCA output; any comparison would be unfair.}

\section{Discussion}
During the course of this analysis, we identified several limitations of the algorithms we tested for solar image segmentation, some of which we have resolved. These include the shape of solar theme distributions, lack of distinct theme signatures, difficulty collecting sufficient expert labeling, and problems specific to operationalizing these algorithms. In the following section we discuss these challenges and, when it was possible, how we resolved them.

\subsection{Non-Gaussianity of distributions}
\begin{figure}
   	\centering
   	\includegraphics[width=\textwidth]{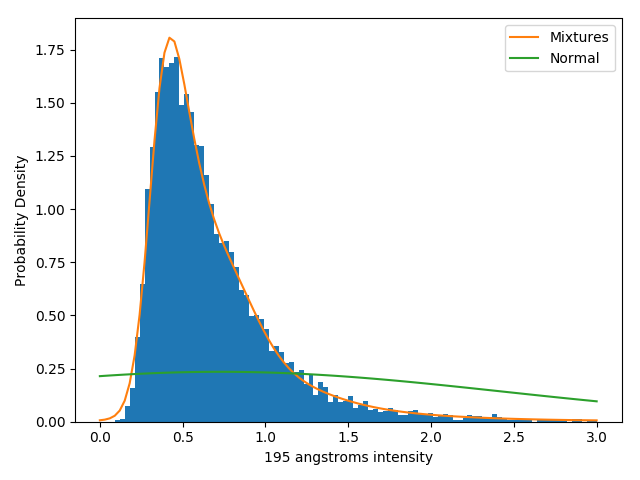}
   	\caption{\textbf{Non-Gaussianity of distributions:} If one assumes a Gaussian distribution (green curve) for themes and directly estimates the classifier parameters, a poor fit to the data results. Using a Gaussian mixture model fitted with the expectation-maximization algorithm, is a better fit to the data (orange curve).}
   	\label{fig:mix}
\end{figure}

\begin{figure}
   	\centering
   	\includegraphics[width=\textwidth]{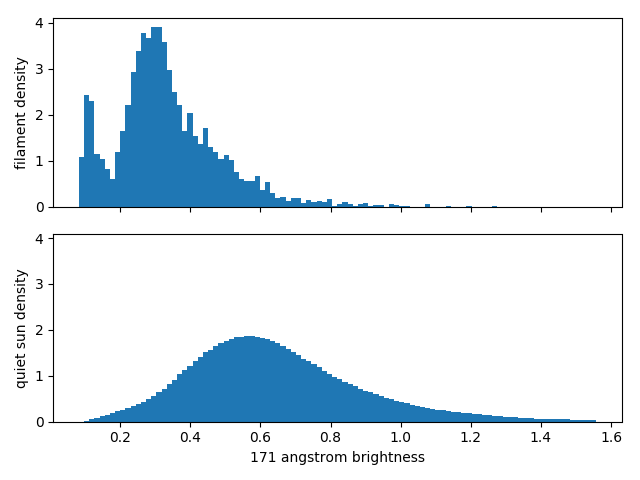}
   	\caption{\textbf{Lack of distinguishing data for filaments:} The 171 ~\AA\ channel separates filaments and quiet Sun the best among EUV. All other channels completely overlap their signatures.}
   	\label{fig:nonunique}
\end{figure}

In the original approach proposed by \citet{rigler+2012}, the distribution of each class could be modeled by a multi-variate Gaussian. While this is a reasonable first-order assumption, it does not appear to be true for real SUVI observations, as some themes have long tails and more asymmetrical distributions. These tails can be important; for example, flares are characterized based on these tails. Furthermore, since \citeauthor{rigler+2012} assumed the distribution was Gaussian, and therefore could not fit long tails, this skewed the means and exaggerated the covariance matrix, resulting in a poor fit (as shown in Figure \ref{fig:mix}). As seen in Figure~\ref{fig:nonunique}, the marginal distributions can be asymmetrical, which implies that they are in fact non-Gaussian distributions. A mixture of multiple Gaussians can fit any distribution. 

\subsection{Lack of distinct theme signatures}\label{sec:los}

Since all classification is based on the spectral properties of a pixel, there is no spatial information included in this analysis. Our inquiry indicates that filaments are not spectrally well separated from quiet sun. This makes it difficult to classify based on spectra alone. We included \halpha to improve the separation, and it does, but found that filament channels in EUV do not always match up with dark filaments in \halpha. This causes further confusion to the classifier. 

Beyond simply lacking sufficiently distinguishing input images, filaments have complicated geometries depending on the line of sight. While training, we noticed that one of our filaments had an especially eccentric spectral signature; it was much brighter than the others. After double checking that it was indeed a filament and not a mislabeled structure, we realized that its proximity to the limb of the Sun meant the dark filament was obscured by bright, diffuse corona in the foreground. Our model assumes every pixel has a single theme, when in reality the three dimensional geometry means we could see both quiet Sun and filament in a pixel. This could lead to more confusion between already hard to distinguish classes. 

\subsection{Training issues}
Developing a database of expert labeled maps is a time consuming process because experts must examine multiple channels to confidently classify regions. In an operational context, training must be continually maintained in order to compensate for changes in instrument performance and the globally changing Sun. Although care must still be taken to obtain high-quality training data, expert classification using our new software is significantly less time consuming than manually drawing features on the Sun. {\color{black} Another approach is to crowd source labeling with citizen scientists in a system like GalaxyZoo, either by asking them to do labeling from scratch or correct proposed boundaries generated from the algorithms. Even when using unsupervised techniques, we need trustworthy labeling for testing.}

We obtained improved classification (versus a single independent annotator) by having several experts label each image and then collaborate. Their complementary expertise and analysis helps avoid misclassification. While one expert might miss many filaments, another expert who studies filaments might identify all and make different mistakes. Using such an approach in operational contexts is strongly recommended: This will help reduce errors in training data and improve the overall performance of any automated solar feature classifier.

Different experts, even those who have similar expertise, might have different opinions on what constitutes a member of the theme. As we show in Figure \ref{fig:ExpertAgreement}, where each expert drew the boundary differed depending on what channels they focused on and on their expertise. This subjectivity creates an inconsistent training database; even with consensus labeling, one group of experts will differ from another group of experts. Each selection could easily overlap with another theme depending on how conservatively the expert labeled. To mitigate this, it would be beneficial to have a much larger pool of experts and a much greater number and variety of images than we used in this study. One anticipated source of labeled maps are the forecasters themselves. Collecting their daily drawings could augment the database considerably. Alternatively, we could utilize rough labelings from experts or citizen scientists and a weakly supervised machine learning approach.

\subsection{Operational concerns} \label{sec:operational}
There are several concerns for running any of these algorithms in an operational setting that have not been explicitly addressed, but are cause for concern and require future work. First, there are times when a channel's image arrives late or contains artifacts due to bad pixels, incomplete data transmission, or spikes due to energetic particle impacts on the detector, resulting in missing or nonsensical values for pixels. We have not explicitly adopted a process to handle this and currently do not create a thematic map at that time. With only one channel missing, it should still be possible to create a thematic map, albeit likely of lesser quality. One could impute a value by taking the observed mean brightness for that pixel location in recent history. This is similar to just using a previous good quality image for that channel. For random forests, there are some more novel approaches to imputing missing data involving proximity searches, (see Section 8 of \cite{ishwaran2008}). 

However, one must be careful to not create new problems with these approaches. In the current incarnation, any errors in the algorithm due to missing data are obvious to a user, as the classifier either outputs a blank thematic map or a severely degraded map with speckles and clearly erroneous classifications. The latter case can occur if one of the high-dynamic-range images only has a short or long exposure image and is not truly high-dynamic-range. This causes the absolute brightness to be wrong and not match any class distribution. An attempt to correct for this using recent pixel values might result in a map that, at first glance looks correct but has hidden biases, e.g. the coronal hole boundaries could be incorrectly defined. Forecasters need to be able to clearly determine when the algorithm is performing as expected, otherwise they will not trust the product and its utility is decreased. 

\begin{figure}
   	\centering
   	\includegraphics[width=\textwidth]{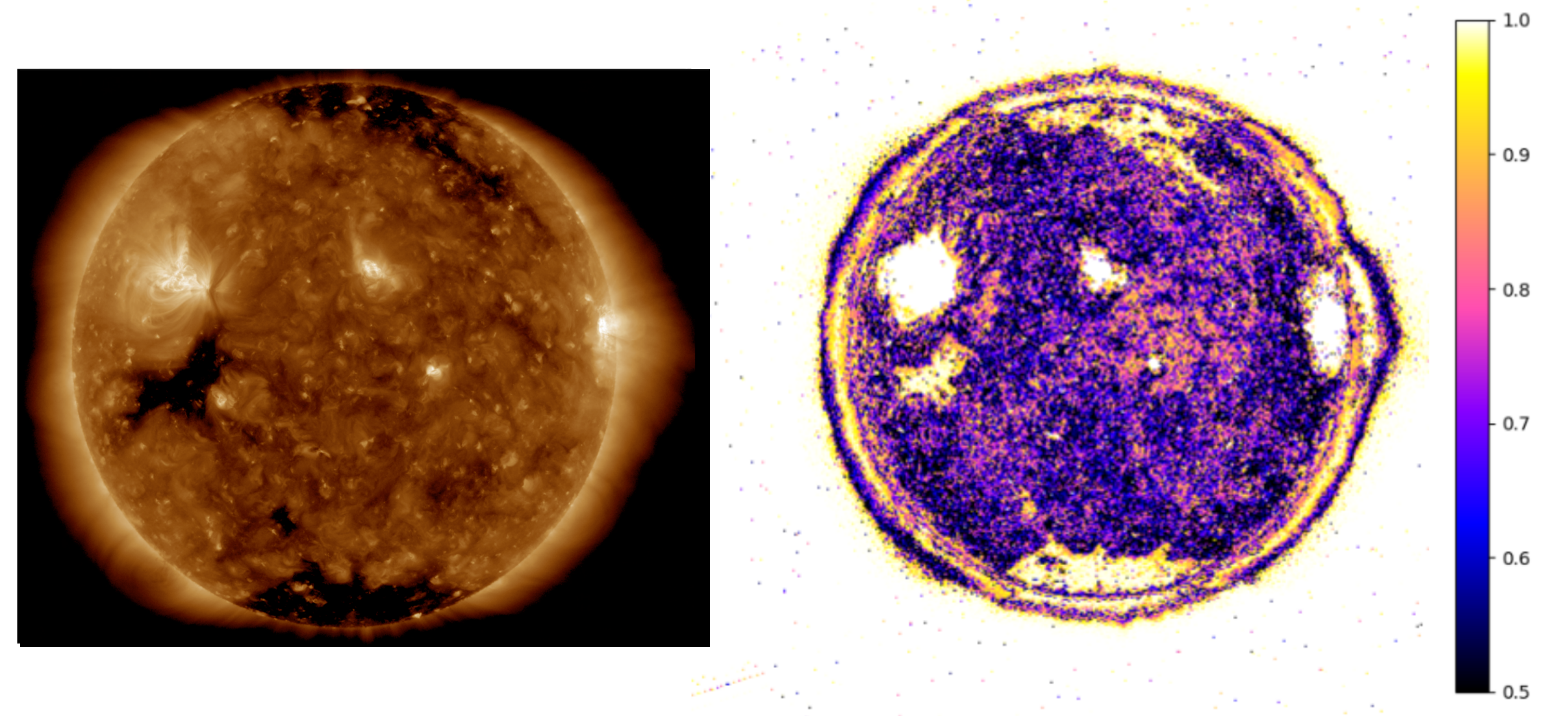}
   	\caption{\textbf{Mixtures Confidence Map for the mixtures approach:} The ``confidence map," (right) compared to a similarly timed SUVI 195~\AA\ observation (left). A high value of 1 corresponding (white) in the confidence map indicates the classifier was certain in the labeling while  values around 0.5 (black) indicates the classifier found two classes nearly equally likely. This is helpful in the cases of filaments and prominences where the signature compared to quiet Sun is not always clear.}
   	\label{fig:confidence}
\end{figure}

We discovered that the thematic map inherently has the possibility to mislead forecasters by reporting a single classification, even in cases where another class was almost as likely for classification. For the mixtures model, we explored creating confidence maps, like in Figure \ref{fig:confidence}. Since the mixtures model reports a likelihood of classification, we report how much that likelihood dominated other classification probabilities: $\text{conf}(i,j) = \frac{\max_{c_m} P(c_m|x_{i,j})}{\sum_{c_m \in C} P(c_m|x_{i,j})}$. For example, if a given pixel has a 50\% likelihood for coronal hole, 5\% likelihood for quiet sun, and 0\% for all other themes, then the reported confidence is $\frac{50}{50 + 5} \approx 0.9$. 
These confidence values do not report absolute confidence, only relative to other classes. Even if the classifier is not confident, on average, the confidence map could indicate a solid performance. 

{\color{black} Two notable ways that the classifier performance can change include if the instrument response is changing or if the training is not well suited for the current solar environment, e.g. the training was done during solar minimum but is being run during solar maximum. The former issue is well documented for AIA's longer wavelength channels and is thought to be caused by satellite propellant contaminating the optical path \citep{Boerner2014}. Analysis of SUVI performance over the course of its lifetime so far suggests this issue is likely experienced in the 284 and 304~\AA\ channels. Without correction, this implies that new training observations will routinely be required as older observations will reflect a different distribution in those channels. However, a real-time correction is presently being implemented in the SUVI data calibration pipeline. Once implemented, degradation is unlikely to have a major impact on the performance of the classifier.

On the other hand, retraining will be required regardless of a degradation correction because as the Sun cycles from solar minimum to maximum, the class distributions will likely change. It is not straightforward to assess how often such a correction might be required. Conditions in the corona can sometimes evolve rapidly, as in, for example, early September 2017 when the Sun rapidly progressed from overall quiet conditions to one of the most active periods of the whole solar cycle over the course of about one week \citep[see, for example,][]{Redmon2018}. In such fast-evolving cases, there is a risk of training data quickly becoming outdated. However, most of the time the Sun's evolution is more gradual, and it is likely that providing updated training perhaps once a month would likely be more than sufficient to accommodate the Sun's evolution.}

\section{Conclusion}
\label{sec:lessons}
We proposed the first random forest model for solar image segmentation and compared it to maximum likelihood approaches using a single Gaussian with a normal distribution assumption and a Gaussian mixtures with a relaxed distribution assumption. The random forest model outperforms the Mixtures or single Gaussian algorithms on SUVI data, due in part to the non-Gaussianity of the theme distributions, and is more robust to noise in images. Furthermore, this investigation concludes that purely spectral based, pixel-by-pixel classification is insufficient for solar image classification as particular themes, most notably filament and quiet sun, are not well differentiated without spatial information. {\color{black} One potential future work is to compare k Nearest neighbors, support vector machines, or other machine learning algorithms to determine the minimal constraints needed to solve this problem, e.g. is the boundary between classes sufficiently described by a linear separator instead of the more complicated random forest.}

This work indicates that future machine learning approaches to solar image segmentation should look at more than just single pixel data but must include spatial information. Instead of labeling pixels, such classifiers {\color{black}should} label regions, which can be combined to produce a similar thematic map product. The necessity of spatial information is especially true if more classes such as sigmoid or emerging flux are to be included. 

One promising approach would be to use convolutional neural networks. However, this requires a large labeled image database, which has proven difficult to obtain. As such, it would be interesting to explore using unsupervised or weakly supervised classification algorithms with many texture filters from computer vision research to complement the raw channel values. Finally, the cyclical, non-stationary dynamics of the Sun present an exciting opportunity for online machine learning algorithms, i.e. algorithms that take examples over time and learn from each new example iteratively. Our algorithm will likely require constant retraining as the Sun emerges from the current minimum to solar maximum and so on, because the underlying theme distributions will likely change. This further highlights the necessity of online algorithms for the next generation of solar image segmentation tools that are able to cope with solar evolution on many different timescales. One potential approach is to use an ensemble of many base classifiers trained on different solar conditions and an online expert switching algorithm \citep{online} to select which base classifier to use at a given time. 

\begin{acknowledgements}
    The authors thank Claire Monteleoni for supporting this research as J. Marcus Hughes's PhD advisor as well as Joonsuk Park for encouraging the project as his undergraduate advisor. We thank Steven Hill of NOAA's Space Weather Prediction Center for helpful advice and encouragement throughout the project. We thank E. Joshua Rigler of the United States Geological Survey for helpful conversations about this work. Support for JMH came from the Boulder Solar Alliance Research Experience for Undergraduate Program funded by NSF grant 1659878. The CIRES contribution to this paper was performed under NOAA cooperative agreement NA17OAR4320101. The editor thanks Martin Reiss and an anonymous referee for their assistance in evaluating this paper.
\end{acknowledgements} 
	
\bibliography{main}
\end{document}